\documentclass[epsf]{elsart}
\def\dfrac#1#2{{\textstyle{#1\over #2}}}
\newcommand{\gev}{\rm{GeV}}
\newcommand{\tev}{\rm{TeV}}
\input epsf
\begin{document}
\begin{frontmatter}
\title{ Impact of the bounds on Higgs mass and $m_W$
on effective theories}
\author[ifuap]{J.L. Diaz-Cruz},
\author[ucr]{J.M. Hern\'andez \thanksref{pds}},
\author[buap]{J.J. Toscano}
\address[ifuap]{Instituto de F\'\i sica, BUAP, 
 Ap. Postal J-48, 72500 Puebla, Pue. M\'exico}
\address[ucr]{Department of Physics, University of California,
 Riverside, EUA}
\address[buap]{Facultad de Ciencias F\'\i sico-Matem\'aticas, BUAP,
      Ap. Postal 1152, 72000 Puebla, Pue. M\'exico }
\thanks[pds]{Permanent address: FCFM, BUAP, Ap. Postal 1152, 72000
Puebla, Pue. M\'exico}

\begin{abstract}
We study the inter-relations that exist between the present experimental
bounds on the Higgs mass, as obtained from radiative corrections to
$m_W$, and the effective parameters, $\alpha_i$ and $\Lambda$. We find
that the SM bounds on $m_H$, arising from a precise determination of the
$W$ mass, can be substantially modified by the presence
of dimension-6 operators which appear in the linear realization of the
effective Lagrangian approach. A Higgs mass as heavy as 700 \gev{} can be
allowed for scales of new physics of the order of 1 \tev. 
\end{abstract}
\end{frontmatter}

\section{Introduction}
 
Some standard model (SM) parameters have been measured with such a high
precision that has allowed to constrain the values of other SM parameters,
or even new physics, through the use of radiative corrections\cite{lepa},
as can be exemplified by the correct agreement between the predicted
top mass and the observed value \cite{topmass}. Finding the Higgs boson
remains as the final step to confirm the theoretical scheme of the SM. The
present lowest experimental bound on the Higgs mass is $m_H>90.4$ \gev{}
\cite{mhexp}, this is a direct search limit. On contrary to the top quark
case, radiative correction are only logarithmically sensitive to the Higgs
mass, and thus it is more difficult to obtain an indirect bound. However,
fits with present data seems to favor a light SM Higgs \cite{altanew}.
Henceforth, it is interesting to ask how this conclusion will change if
one goes beyond the SM.

The framework of effective Lagrangians, as a mean to parametrize physics
beyond the SM in a model independent manner, has been used extensively
recently \cite{effecl}. Within this approach, the effective lagrangian is
constructed by assuming that the virtual effects of new physics modify the
SM interactions, and these effects are parametrized by a series of
higher-dimensional nonrenor\-ma\-lizable operators written in terms of the
SM fields.  The effective linear Lagrangian can be expanded as follows:
\begin{equation}
{\mathcal L}_{\mathrm{eff}}=  {\mathcal L}_{\mathrm{SM}} 
+ \sum_{i,n} {\frac{\alpha_i}{\Lambda^n} } O^i_n
\end{equation}
where $\mathcal{L}_{\mathrm{SM}}$ denotes the SM renormalizable
lagrangian. The terms $O^i_n$ are $SU(3) \times SU(2)_L \times U(1)_Y$
invariant operators. $\Lambda$ is the onset scale where the appearance of
new physics will happen. The parameters $\alpha_i$ are unknown in this
framework, although "calculable" within a specific {\it full theory}
\cite{ital}. This fact was used in Ref. \cite{effecc} to show that, in a
weakly coupled full theory, a hierarchy between operators arises by
analyzing the order of perturbation theory at which each operator could be
generated {\it e.g.} by integrating the heavy degrees of freedom.  Some
operators can be generated at tree-level, and it is natural to assume that
their coefficients will be suppressed only by products of coupling
constants; whereas the ones that can be generated at the 1-loop level, or
higher, will be also suppressed by the typical $1/16\pi^2$ loop factors.
This allows us to focus on the most important effects the high-energy
theories could induce, namely those coming from tree-level generated
dimension-six operators.

In this letter we address two related questions. First, we study how the
effective lagrangian affects the determination of the W boson mass,
and how this could affect the bounds on the Higgs mass.  The second item
under consideration will be to re-examine the effects on Higgs-vector 
boson production at hadron colliders by using the results obtained from
the first part. 

\section{The SM $W$ mass}
We shall use the results of Ref. \cite{degrassi,coefi}, which parametrize
the bulk of the radiative corrections  to  $W$ mass through the
following expression:
\begin{equation}
m_W=m_W^o \ [1+ d_1 ln (\frac{M_h}{100})+d_2 C_{em}+
    d_3 C_{top} +d_4 C_{as} + d_5 ln^2(\frac{M_h}{100}) ],
\end{equation}
where the coefficients $d_i$ are given in table 2 of Ref.
\cite{coefi}, they incorporate the full 1-loop effects, and
some dominant 2-loop corrections. The factors $C_i$ are
given by:
\begin{eqnarray}
C_{em}&=&\frac{\Delta \alpha_h}{0.0280} -1, \nonumber \\
C_{top}&=&(\frac{m_t}{175 GeV})^2 -1,          \nonumber \\
C_{as}&=&\frac{\alpha_s(M_Z)}{0.118} -1,
\end{eqnarray}
and they measure the dependence on the fine structure 
constant, top mass and strong coupling constant, respectively. 
The reference $W$ mass, $m_W^o=80.383$ \gev, is obtained with the
following values: $mt=175$ \gev, $\alpha_s=.118$, 
$\Delta \alpha=0.0280$, and $m_H=100$ \gev.
By using eqs. (2-3), one finds a bound on the SM Higgs mass,
$m_t=176 \pm 2$ \gev, $\Delta \alpha=.0280$
and $\alpha_s=.118$, of $170< m_H<330$ \gev.
This result agree with several other studies \cite{lepa} which
anticipates the existence of a light Higgs boson. 

Including the effects of new physics will modify the W mass value. That 
effect should be combined with the previous SM corrections in order to
determine to what extent new physics could change 
the bounds obtained for the Higgs mass.

\section{Modification to the $W$ mass}
The complete set of effective operators is large but
the analysis simplifies because loop-level dimension six operators and
tree-level dimension eight ones generate subdominant effects with respect
to tree-level generated dimension six effective operators \footnote{This
is not the case when tree-level dimension six operators do not contribute
\cite{yomi,mythes}.}. The effective contributions to the input parameters
in the formulas (2-3), besides $m_W$, can be show to dissappear after 
suitable redefinitions \cite{mythes}. Just two operators contribute:
 \begin{eqnarray}
O_{\phi}^{(1)}&=&(\phi ^{\dagger }\phi ) (D^\mu \phi ^{\dagger }
		D_\mu \phi), \nonumber \\
O_{\phi}^{(3)}&=&(\phi ^{\dagger } D_{\mu} \phi) (D^{\mu} 
                \phi)^{\dagger } \phi. 
\end{eqnarray}
The notation is as usual, $\phi$ denotes the Higgs doublet,
and $D_{\mu}$ is the usual covariant derivative. 
An interesting characteristic of these operators is that 
only $O_{\phi}^{(1)}$ gives contribution to $m_W$; although both 
operators modify the coefficients of the vertices $HVV$ they leave 
intact its Lorentz structure. 
This approach is equivalent to the most usual one used in the literature
in which more operators are allowed but they are constrained in a
one-per-one basis, so that "unnatural" cancellations are not allowed.

We shall now include the contribution to $m_W$ arising from the effective
operators of eq. (4) into eqs. (2). The formulae for the $W$ mass
becomes:
\begin{equation}
m_W |_{\mathrm{eff}} \ = \ m_W |_{\mathrm{SM}} \ 
(1+\frac{1}{4} \alpha_{\phi}^{(1)} (\frac{v}{\Lambda})^2)
\end{equation}
where $m_W |_{\mathrm{SM}}$ corresponds to the $W$ mass as defined 
in eq. (2), and the term in the parentheses arises from
the effective Lagrangian. 

To study the inter-relations between the Higgs mass and
$\alpha_{\phi}^{(1)}$ and $\Lambda$, we must set the allowed $m_W-m_H$
region. We are going to use the future expected
minimum uncertainty for the W mass: $\Delta m_W=\pm.01$ \gev{}
for a nominal central value of $m_W=80.33$ \gev,
and expand $m_H$ between the recent experimental bound, 
$m_H \ge 90$ \gev{} \cite{mhexp}, and the perturbative limit, 
$m_H \le 700$ \gev (Fig. 1). We take an optimum scenarios also for
the top quark mass ($ m_t=176 \pm 2$ \gev).
$\alpha$ and $\Lambda$ should take values such that the resulting masses
must satisfy the above constraints. Since we assumed the 
effective Lagrangian is derived from a weakly coupled 
theory, it is reasonable to impose\footnote{We will choose 
the $\alpha_i$ signs that give the maximum and minimum values 
for the quantities of  interest.} $|\alpha_{\phi}^{(1)}| \le 1$
, while $\Lambda$ is set
to values greater or equal than 1 \tev. This scale is a conventional one,
it can be justified in some specific high-energy models. 
It turns out that the case when
$|\alpha_{\phi}^{(1)}|=1$ and $\Lambda=1$ \tev, simultaneously,
is already excluded by present data.
\begin{figure}
\begin{center}
\epsfysize=8cm  \epsfbox{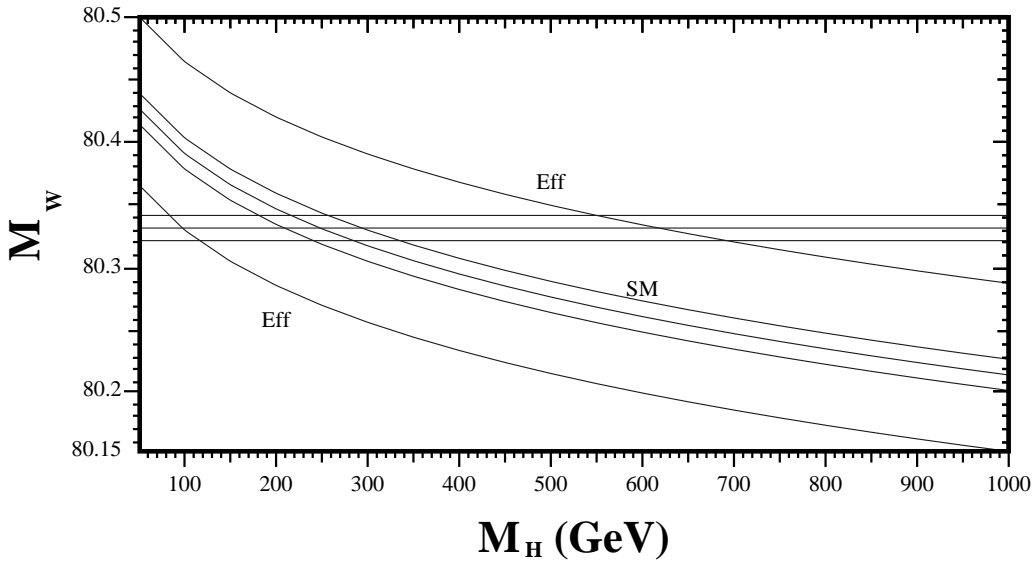}
\end{center}
\caption{$m_W$ versus $m_H$ for $m_t=176 \pm 2$ \gev, both in the Standard
Model (SM) and in the maximum effective contributions (Eff).}
\label{uno}
\end{figure}
\begin{figure}
\begin{center}
\epsfysize=6cm \epsfbox{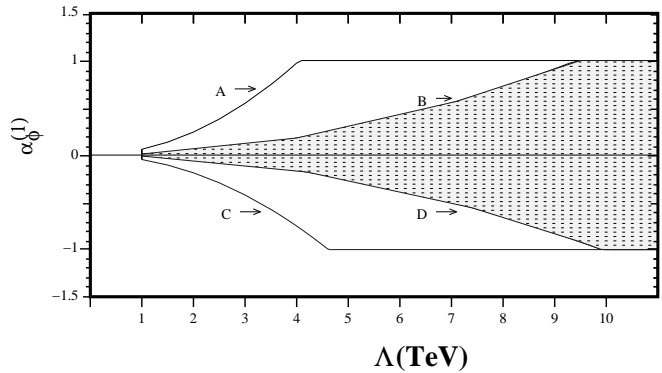}
\end{center}
\caption{Constrained region for $\alpha_{\phi}^{(1)}$ and $\Lambda$.
$m_t=176$ \gev.}
\label{dos}
\end{figure}
Figure \ref{dos} shows the level curves in the $\alpha-\Lambda$ 
plane according with the bounds discussed before for $m_W$ and $m_H$.
The allowed region of parameters corresponds to the area located to the
right of curves A, B, C, D. Allowing for an enlargement of the allowed
Higgs mass  bound from $170 <m_H<330$ \gev{} in the SM to $90 < m_H < 700$
\gev{} into the effective Lagrangian case. The curve A is obtained
by taking $m_H=700$ \gev{} and $m_W=80.32$ \gev, while for the curve B we
use $m_H=90$ \gev{} and $m_W=80.34$ \gev.  The shadowed area between the
curves (BD) marks the parameters region where no new physics effects can
be disentangled from the SM uncertainties. These results were obtained by
considering $m_t=176$ \gev; it is  found that there are not substantial
changes by adding the top mass uncertainty to the effective contributions. 
It is also found that the allowed ranges for $\Lambda$ and
$\alpha_{\phi}^{(1)}$ are as follow: for $\Lambda=1$ \tev, 
$.011< \alpha_{\phi}^{(1)} <.060$ and 
$-.01010< \alpha_{\phi}^{(1)} < -.01019$. This correspond to Higgs masses
between the perturbative limit and upper SM bound in the first interval, 
and between the lower SM bound (obtained from radiative corrections to
$m_W$) and the experimental limit for the second interval. In the
complementary case, i.e. taking $\alpha_{\phi}^{(1)}$ equal to its maximum
value, it is found that  $4.1< \Lambda < 9.5$ \tev, and $4.64 < \Lambda <
10$ \tev with same considerations for the Higgs mass as in the first case.
Then, independiently of the  value $\alpha_{\phi}^{(1)}$  effects arising
from a scale beyond of 10 \tev can not be disentangled from SM top
uncertainties.

\section{Associated $W(Z)$ and $H$ production}
\begin{figure}[!h]
\begin{center}
\epsfysize=6cm \epsfbox{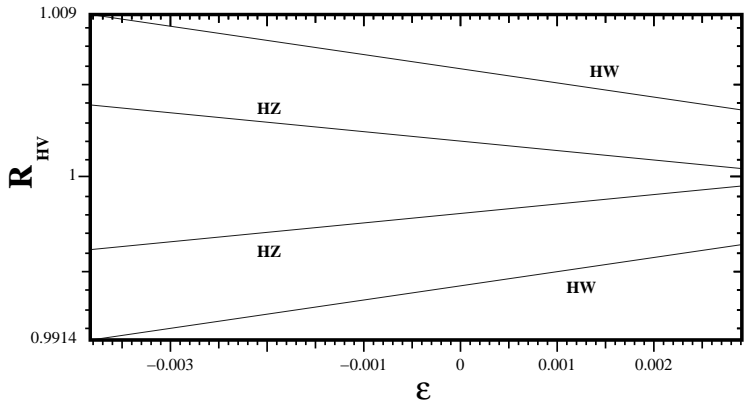}
\end{center}
\caption{Ratio of the effective cross section to the SM one for the
associated $HW,HZ$ production. It is assumed that $\alpha_{\phi}^{(3)}$
behaves in the same way as $\alpha_{\phi}^{(1)}$ does.} 
\label{tres}
\end{figure}

In this paper we also re-examined accordingly the modifications to the 
mechanism of associated production of Higgs boson with a 
vector particle ($W,Z$), due to the effective Lagrangian updating the
results obtained in Ref. \cite{lorent}. The corresponding Lagrangian
to be used is  :
\begin{equation}
{\mathcal L}_{HVV} = \frac{m_Z}{2} (1 + f_1) H Z_{\mu} Z^{\mu} 
+g m_W (1+f_2) H W^+_{\mu}W^{-\mu},
\end{equation}
where the parameter $f_i$ are functions of 
$\epsilon_j=\alpha_j (\frac{v}{\Lambda})^2$
given as follows: 
\begin{equation}
f_1=\dfrac{1}{2}(\epsilon_{\phi}^{(1)} 
+\epsilon_{\phi}^{(3)} ), \hspace{2cm}
f_2=\dfrac{3}{4} (2\epsilon_{\phi}^{(1)} 
    -\epsilon_{\phi}^{(3)} ).
\end{equation}
The ratio of the effective cross-section 
to the SM one for the processes $p\bar{p} \to H+V$ have been evaluated. 
The parton  convolution part is factored out, and only remains the ratio
of partonic  cross-sections, thus the result is valid for both FNAL and
LHC. The expressions for the cross-sections ratios are:
\begin{eqnarray}
R_{HW}&=&\frac{\sigma_{\mathrm{eff}}(p\bar{p} \to H+W)}
             {\sigma_{\mathrm{SM}}( p\bar{p} \to H +W)} \nonumber \\
       &=& (1+f_2)^2 	
\end{eqnarray}
\begin{eqnarray}
R_{HZ}&=&\frac{\sigma_{\mathrm{eff}}( p\bar{p} \to H+Z)}
              {\sigma_{\mathrm{SM}}( p\bar{p} \to H +Z)} \nonumber \\
       &=& (1 + f_1)^2
\end{eqnarray}
For the operators under consideration, the cross-sections 
ratio is independent of the Higgs mass, and as
the best values, it is found that the cross-section is only
slightly modified. The behavior of the cross-section ratios
are shown in fig. \ref{tres}, for typical values of $\alpha$
and $\Lambda$ as found in section 4. We consider the same 
estimate for $\alpha_{\phi}^{(3)}$ as we got for $\alpha_{\phi}^{(1)}$.
As it can seen from figure \ref{tres},
these two processes result almost insensible to new physics
effects arising from the dimension-six operators $O_{\phi}^{(1,3)}$,
since their effects is of the order $10^{-3}$. 
The corresponding contributions that arises from
 the effective operators that were neglected are, at least, 2 
orders of magnitude below that ones that are considered here. Of course,
a more detailed study is needed in order to include the modifications
to the expected number of events for discovery as a function of $m_H$,
and that case is under study.

\section{Conclusions}  
We have studied the modifications that new physics imply for 
the bound on the Higgs mass that is obtained from radiative corrections
to electroweak observables, within the context of effective
lagrangians. We found that the SM bound 
$170 \le m_H \le 330$ \gev{} that is obtained from a precise determination
of the $W$ mass, can be substantially modified by the presence of
dimension-6 operators that arise in the linear realization 
of the effective Lagrangian approach. A Higgs mass as heavy as
700 \gev{} is allowed for scales of new physics of the order of
1 \tev, with a corresponding value for $|\alpha_{\phi}^{(1)}|$ of
the order of $10^{-2}$. Aswell we found that even for
$|\alpha_{\phi}^{(1)}|=1$, new physics effects arising from scales
$\Lambda > 10$ \tev{} can not be separate from the uncertainties on
the top quark mass, in an optimum scenarios for the
observables considered here. Those results give us the landmark for the
decoupling limits for both $\alpha_i$ and $\Lambda$.
Accordingly, it is found that such operators do not
produce a significant modification for the present (FNAL) or future (LHC)
studies for the associated production mechanism $p\bar{p} \to H+V$.

\ack 
We acknowledge financial support from CONACYT and SNI (MEXICO). We also
acknowledge to M.A. P\'erez for discussions.

\end{document}